\documentclass[12pt]{article}
\usepackage{epsfig}
\usepackage{graphicx}
\usepackage{latexsym}
\usepackage{ifpdf}
\ifpdf
\usepackage{textcomp}
\usepackage{amssymb}

\usepackage{amsfonts,amsthm,amstext,amscd}
\usepackage{slashed}
\usepackage{amsfonts,amssymb,amsthm,amstext,amscd}

\usepackage{amsmath}

\newcommand{\beqa}{\begin{eqnarray}}
\newcommand{\eeqa}{\end{eqnarray}}

\newcommand{\be}{\begin{equation}}
\newcommand{\ee}{\end{equation}}
\newcommand{\beq}{\begin{equation}}
\newcommand{\eeq}{\end{equation}}
\newcommand{\bea}{\begin{eqnarray}}
\newcommand{\eea}{\end{eqnarray}}

\newcommand{\bear}{\begin{eqnarray}}
\newcommand{\eear}{\end{eqnarray}}
\begin{document}
\baselineskip=15.5pt
\pagestyle{plain}
\setcounter{page}{1}

\def\r{\rho}
\def\CC{{\mathchoice
{\rm C\mkern-8mu\vrule height1.45ex depth-.05ex
width.05em\mkern9mu\kern-.05em}
{\rm C\mkern-8mu\vrule height1.45ex depth-.05ex
width.05em\mkern9mu\kern-.05em}
{\rm C\mkern-8mu\vrule height1ex depth-.07ex
width.035em\mkern9mu\kern-.035em}
{\rm C\mkern-8mu\vrule height.65ex depth-.1ex
width.025em\mkern8mu\kern-.025em}}}

\newfont{\namefont}{cmr10}
\newfont{\addfont}{cmti7 scaled 1440}
\newfont{\boldmathfont}{cmbx10}
\newfont{\headfontb}{cmbx10 scaled 1728}
\renewcommand{\theequation}{{\rm\thesection.\arabic{equation}}}

\par\hfill{IFUP-TH-2015-05}
\vspace{1cm}

\begin{center}
{\huge{\bf Notes on Theta Dependence in Holographic Yang-Mills
}}
\end{center}

\vskip 10pt

\begin{center}
{\large Francesco Bigazzi$^{a}$, Aldo L. Cotrone$^{b}$ and Roberto Sisca$^{c}$}
\end{center}

\vskip 10pt
\begin{center}
\textit{$^a$ INFN, Sezione di Pisa; Largo B. Pontecorvo 3, I-56127 Pisa, Italy.}\\
\textit{$^b$ Dipartimento di Fisica e Astronomia, Universit\`a di
Firenze and INFN, Sezione di Firenze; Via G. Sansone 1, I-50019 Sesto Fiorentino
(Firenze), Italy.}\\
\textit{$^c$ Universit\`a di Pisa - Dipartimento di Fisica ``E. Fermi";
Largo Bruno Pontecorvo 3, I-56127 Pisa, Italy.}\\

\vskip 10pt
{\small fbigazzi@pi.infn.it, cotrone@fi.infn.it, r.sisca@studenti.unipi.it}
\end{center}

\vspace{25pt}

\begin{center}
 \textbf{Abstract}
\end{center}
  
\noindent
Effects of the $\theta$ parameter are studied in Witten's model of holographic 4d Yang-Mills, where $\theta$ is the coefficient of the CP-breaking topological term. First, the gravity background, including the full backreaction of the RR form dual to the $\theta$ parameter, is revisited.
Then, a number of observables are computed holographically: the ground-state energy density, the string tension, the 't Hooft loop, the light scalar glueball mass, the baryon mass scale, the critical temperature for deconfinement - and thus the whole $(T,\theta)$ phase diagram -  and the entanglement entropy. A simple rule is provided to derive the $\theta$ corrections to (at least) all the CP-neutral observables of the model. Some of the observables we consider can and have been in fact studied in pure 4d Yang-Mills on the lattice. In that framework the results, obtained in the small $\theta$ regime, are given up to very few powers of $\theta^2$. The corresponding holographic results agree qualitatively with available lattice data and signal an overall mass scale reduction by $\theta$. Moreover, being exact in $\theta$, they provide a benchmark for higher order corrections in Yang-Mills.

\newpage

\section{Introduction and results}
\setcounter{equation}{0}
The top-down prototype of holographic Yang-Mills, due to Witten \cite{witten}, is based on the background generated by $D4$-branes wrapped on a circle with appropriate boundary conditions. The field theory dual is a non supersymmetric confining Yang-Mills coupled to massive ``Kaluza-Klein" (KK) matter in the adjoint representation. In this note we are interested in studying the effects of the Yang-Mills $\theta$ angle - the coefficient of the CP-breaking topological term in the Lagrangian - in this model. 

There exist interesting studies of the $\theta$ dependence of some observables in lattice Yang-Mills, see for example the excellent review \cite{vicari} and references therein.
As usual, these studies can be performed in principle at any number of colors $N_c$ and for the lattice discretization of actual pure Yang-Mills. 
On the other hand they are challenging, due to a sign problem related to the $\theta$ term. 
For this reason, lattice results are limited to small values of $\theta$ (mainly to the first few non trivial orders in the $\theta^2$ expansion around $\theta=0$), as they are either analytically continued from imaginary $\theta$ values or obtained by computing certain correlators at $\theta=0$.

In the holographic model there is no such limitation. When $\theta$ is very small, one can treat the corresponding (1-form Ramond-Ramond) field in the dual gravity background as a probe \cite{Witten:1998uka}. It is not difficult, however, to account for the full backreaction of that field and so to consider finite generic values of $\theta$. The corresponding gravity solution has been found in \cite{pasquinucci,dubovsky}. This is what we will focus on in this note.  

In the holographic framework, the main limitations arise from the fact that the classical gravity solution, dual to the field theory, is reliable only in the planar $N_c\gg1$ limit at ``strong coupling" $\lambda_4\gg1$, where $\lambda_4$, which can be viewed as the Yang-Mills 't Hooft coupling at the Kaluza-Klein mass scale $M_{KK}$, is a parameter which actually measures how much the spurious KK matter fields are decoupled from the Yang-Mills theory. When $\lambda_4$ is large the decoupling cannot be achieved. Despite this limit, Witten's model realizes in a very simple way all the crucial expected IR features of the pure Yang-Mills theory. In this note we want to provide some novel evidence for this to be true also at finite $\theta$-angle. As we will see, also because the $\theta$ term enters in Witten's Yang-Mills theory in the same way it enters in pure Yang-Mills, the holographic approach is able to capture the same qualitative trends of the topological effects expected in pure Yang-Mills theory. This justifies the comparison we will make with lattice results on the way. 

At large $N_c$ and $\lambda_4$, the effects of the $\theta$ angle turn out to be actually weighed by the combination $\Theta \sim \lambda_4\, \theta/N_c$. In order for non-trivial effects due to the $\theta$ angle to be considered, the limits have to be taken in such a way that $\Theta$ remains finite. 

After having presented the gravity background in Section \ref{bac} (see also Appendix \ref{appA} for a re-derivation), in Section \ref{obs} we enroll in the study of the $\theta$ dependence of a number of relevant observables in the dual gauge theory.
Although the theory is not precisely pure Yang-Mills and it is studied in the planar limit, it allows to derive the \emph{exact} $\theta$ dependence of the observables, providing a complementary view with respect to lattice Yang-Mills.

Some of the observables we calculate have been studied in lattice Yang-Mills: the fundamental string tension, the light scalar glueball mass, and the critical temperature $T_c$ for deconfinement have all been computed to order $\theta^2$ \cite{Del Debbio:2006df,massimo}. In all these cases, at the same order, the observables in the holographic model agree qualitatively\footnote{We do not report on a comprehensive quantitative comparison of the holographic model with the lattice data. The latter are still affected by large statistical errors in the large $N_c$ limit, and the two approaches currently seem to present significant discrepancies.} with the lattice results - namely the $\theta^2$ corrections to the $\theta=0$ values are all \emph{negative}: the $\theta$ term reduces the mass scales.

The ground-state energy density, whose ${\cal O}(\theta^2)$ coefficient gives the topological susceptibility, has also been computed on the lattice to order $\theta^4$ (the ${\cal O}(\theta^6)$ coefficient has been considered e.g. in \cite{t41,t42}: unfortunately it is not possible to determine its sign yet, as the errors are still very large). Also in this case, holographic and lattice data qualitatively agree.

Interestingly enough, in the holographic model the above $\theta$-corrections are just the first non trivial terms of (powers of) geometric series in $\Theta^2$. More precisely, they come from the expansion of functions like $(1+\Theta^2)^{-a}, a>0$. As such, the holographic model \emph{predicts} that the ${\cal O}(\theta^4)$ corrections to, say, the string tension and the glueball mass, will be of positive sign. Analogously, the ${\cal O}(\theta^6)$ coefficient in the ground-state energy density (which has an overall negative sign) is predicted to be positive. It would be very interesting to improve the lattice precision in order to check whether these predictions are actually realized in pure Yang-Mills.

As another notable feature, the holographic model precisely accounts for the expected invariance of the gauge theory observables under $\theta\rightarrow \theta+2\pi$ shifts. For instance, the expected (see e.g. \cite{massimo}) periodic structure of the $(T,\theta)$ phase diagram with triple points where first-order phase transition lines meet, is explicitly realized.

In the present work, we also consider the $\theta$ dependence of quantities for which there are no lattice results yet, namely the 't Hooft loop (from which we deduce an explicit realization of the so-called ``oblique confinement"), the mass scale of baryons,\footnote{In \cite{cai} baryons have been investigated in the related solution in \cite{pasquinucci}.} which follows the same pattern as the observables described above, and the entanglement entropy.
The latter is studied for two different geometries, the ``slab'' and the ball.
In both cases there is a phase transition as the dimension of the entangling region is varied, between a ``IR configuration'' and a ``UV configuration'' \cite{klebanov,wen}.
While the former scales like $(1+\Theta^2)^{-a}$, the latter is insensitive to $\Theta$, consistently with the fact that the $\theta$ dependence is a IR phenomenon.
The critical length for the transition grows as $(1+\Theta^2)^{1/2}$.

All of the $\Theta$ corrections to the observables in the holographic model are extremely simple: they are given by multiplying (factors of) the $\Theta=0$ result by powers of $(1+\Theta^2)$.
It is actually a-posteriori trivial to guess which power of $(1+\Theta^2)$ appears in each observable.
In fact, the model has two distinct mass scales: the Kaluza-Klein scale $M_{KK}$ (which is also the glueball mass scale at $\Theta=0$) and the string tension $T_s$.
Their ratio is determined by the parameter $\lambda_4 \sim T_s/M^2_{KK}$.
As it turns out, it is sufficient to determine how the two quantities $M_{KK}$ and $\lambda_4$ (or equivalently $T_s$) scale with $\Theta$ to have the scaling of all the observables.
To be specific, it is sufficient to include a factor of $(1+\Theta^2)^{-1/2}$ for each power of $M_{KK}$ and a factor of $(1+\Theta^2)^{-1}$ for each power of $\lambda_4$ appearing in a given observable.

Although we have no counter-example to this rule, we cannot exclude that observables which are sensible to CP parity evade it, due to mixing effects.\footnote{The mesonic spectra in \cite{cai2} could be of this kind.}
Apart from this caveat, the rule above would allow to write down the $\Theta$ corrections to at least \emph{all} the CP-neutral observables of the model, once their $M_{KK}$ and $\lambda_4$ factors are known at $\Theta=0$,\footnote{This rule would suggest in turn that observables scaling, at $\Theta=0$, as powers of  $\lambda_4 M_{KK}^{-2}$, are not corrected by $\theta$.}
without performing any calculation! This is ultimately due to the simple origin of the $\theta$ deformation of the background in 11d, where it is basically a twist of a two-cycle \cite{Gross:1998gk,dubovsky}.
It would be extremely interesting to find a similar pattern on the lattice, at least for some classes of observables. 

As a final aside comment, let us recall that chiral massless fermions, transforming in the fundamental of the gauge group, can be introduced in Witten's model by adding $D8$-branes, giving the prototype ``holographic QCD'' model of Sakai and Sugimoto \cite{ss}. In that setup, the axial $U(1)_A$ anomaly and the Witten-Veneziano mechanism relating the mass of the $\eta'$ meson with the topological susceptibility of the unflavored theory are precisely realized. At the same time, consistently with what happens in massless QCD, a non zero $\theta$ term in the Sakai-Sugimoto model can be rotated away by a chiral rotation of the fermions. As a result the topological susceptibility of the flavored theory is zero. In order to see the effects of the topological term in this QCD-like setup, one needs to switch to the case where the fundamental matter fields are massive. We hope to come back to these issues in the future.  

\section{Theta dependence in Holographic Yang-Mills}\label{bac}
\setcounter{equation}{0}
The Yang-Mills Euclidean Lagrangian at finite $\theta$ is given by\footnote{We use the standard normalization conventions ${\rm Tr} [t_a t_b] = (1/2)\delta_{ab}$ for the $SU(N_c)$ generators $t_a$. Notice moreover that ${\tilde F}^{a}_{\mu \nu} = (1/2) \epsilon_{\mu\nu\rho\sigma}F^{a\rho\sigma}$. In Minkowski, the Lagrangian has a minus sign on the first term and a plus sign, replacing the imaginary factor $-i$, on the second term.}
\be
{\cal L} = \frac{N_c}{2\lambda}\left[{\rm Tr} F^2 -i \frac{\lambda}{8\pi^2}\frac{\theta}{N_c}{\rm Tr} F{\tilde F}\right]\,,
\label{lagthetaE}
\ee
where $\lambda = g^2 N_c$ is the 't Hooft coupling. The CP breaking $\theta$ parameter multiplies the topological charge density and as such it behaves like an angle: the physics has to be invariant under shifts $\theta\rightarrow \theta+2\pi$. At the same time, as it is evident from the previous expression, in the large $N_c$ limit the theory has a non-trivial $\theta$-dependence (as required, e.g. by the large $N_c$ solution of the $U(1)_A$ problem, i.e. why the $\eta'$ mass is much higher then those of the meson octet) only if $\theta/N_c$ is held fixed. Observables, like the ground-state energy density, will thus be functions of $\theta/N_c$.

As it was shown in \cite{witten80}, a way to reconcile the periodicity requirement and the large $N_c$ scaling is to allow for the vacuum energy density of the theory to be a multi-branched function. Possible vacua, labeled by an integer $k$, become stable but non-degenerate at $N_c=\infty$ and the ground-state energy, for a given value of $\theta$, is obtained by minimizing, with respect to $k$, a function of $(\theta+2\pi k)/N_c$. The resulting expression, periodic in $\theta$, is not regular everywhere: at 
certain values of $\theta$ it accounts for a jump between different branches.

In the holographic Yang-Mills model, this expected behavior is explicitly realized \cite{Witten:1998uka}. 

Witten's original background \cite{witten} is sourced by $N_c\gg1$ $D4$-branes (of type IIA string theory) wrapped on a circle $S_{x_4}$, of length $2\pi M_{KK}^{-1}$, where periodic (resp. anti-periodic) boundary conditions on scalars (resp. fermions) are imposed. In such a way, in the deep IR (i.e. at energies $E\ll M_{KK}$) the original $4+1$ dimensional theory on the $D4$-branes, reduces to pure non-supersymmetric $SU(N_c)$ Yang-Mills in $3+1$ dimensions. In fact, all the matter fields, transforming in the adjoint, get masses of the order of $M_{KK}$.

The Yang-Mills $\theta$-angle is holographically related to the integral of the RR one-form $C_1$ over  the circle $S_{x_4}$. In \cite{Witten:1998uka} this field was taken as a probe of the original $D4$-brane background. This approximation is fine if one is interested in keeping $\theta/N_c$ very small and getting only the leading order corrections in this parameter. 

In this note we want to consider the full backreaction of the $\theta$ angle on the background, with the aim to explore at deep how the IR Yang-Mills physics is affected by it. 

The relevant type IIA gravity action, in string frame, reads
\be
S = \frac{1}{2k_0^2}\int d^{10}x\sqrt{-g}\left[e^{-2\phi}\left({\cal R}+4(\partial\phi)^2\right) -\frac12 |F_4|^2 -\frac12 |F_2|^2\right]\,.
\label{action1}
\ee 
Here $2k_0^2=(2\pi)^7 l_s^8$ where $l_s\equiv\sqrt{\alpha'}$ is the string length, $F_4 = dC_3$ is the RR four-form which is magnetically sourced by the $N_c$ $D4$-branes, $F_2=dC_1$, and $\phi$ is the dilaton.

The $\theta$-backreacted background \cite{pasquinucci,dubovsky} (see appendix \ref{appA} for a detailed re-derivation based on the 1d reduced action) is a solution of the equations of motion derived from the action above. The string frame metric reads
\be
\label{metricu}
ds_{10}^2 =\left(\frac{u}{R}\right)^{3/2}\left[\sqrt{H_0}\, dx_{\mu}dx^{\mu} + \frac{f}{\sqrt{H_0}}\, dx_4^2\right]+ \left(\frac{R}{u}\right)^{3/2}\sqrt{H_0}\left[\frac{du^2}{f} + u^2 d\Omega_4^2\right]\,,
\ee
where
\be
f = 1-\frac{u_0^3}{u^3}\,,\qquad H_0 = 1-\frac{u_0^3}{u^3}\frac{\Theta^2}{1+\Theta^2}\,.
\ee
The background also includes a running dilaton, a RR one-form and a four-form field strength given by
\be\label{dilatonu}
e^{\phi} = g_s \left(\frac{u}{R}\right)^{3/4} H_0^{3/4}\,,\qquad C_1 = \frac{\Theta}{g_s} \frac{f}{H_0} dx^4\,,\qquad F_4 = 3 R^3 \omega_4\,,
\ee
with the flux quantization condition fixing the value of $R$ as
\be
\int_{S^4} F_4 = 8\pi^3 l_s^3 g_s N_c\,, \quad R= (\pi g_s N_c)^{1/3} l_s\,.
\label{f42}
\ee 
Here $\Theta$ is a parameter which, as it will be clear in a moment, is proportional to $\theta$. The original background at zero $\theta$ angle, found in \cite{witten}, can be readily obtained from the one above setting $\Theta=0$. Let us recall that in the expression above $\mu = 0,1,2,3$ are the $1+3$ Minkowski directions where the Yang-Mills theory is defined, $d\Omega_4^2$ is the metric of a $S^4$ of radius one, $u$ is the transverse radial coordinate $u\in [u_0,\infty)$ - holographically mapped to the RG scale of the dual field theory - $x_4$ is the compact coordinate of length $2\pi M^{-1}_{KK}$ and $R$ is a curvature radius. The isometry group of $S^4$ is mapped into a global $SO(5)$ symmetry group in the dual field theory, which acts non-trivially on the Kaluza-Klein massive modes (signaling that these are, in fact, not decoupled in the limit we are considering). Finally $g_s$ is the string coupling and $\omega_4$ is the volume form of the transverse $S^4$, of volume $V_{S^4}=8\pi^2/3$. 

Notice also that, crucially, the $S_{x_4}$ circle shrinks to zero size when $u=u_0$. Absence of conical singularities at $u=u_0$ is guaranteed if 
\be\label{mkk}
u_0 = \frac{4 R^3}{9}M_{KK}^2\frac{1}{1+\Theta^2}\,,
\ee
which thus implies that the $(x_4,u)$ subspace has the topology of a disk. 
 
Reliability of the background requires $e^{\phi}$ to be small: when this condition is violated we should better make use of the 11d (``M-theory") completion of the model. As it was shown in \cite{witten}, in fact, the 10d solution can be obtained starting from an $AdS_7\times S^4$ planar black hole solution of 11d supergravity, reducing it on the M-theory circle and performing a double Wick rotation. We will make use of this picture in the following.

As it will be reviewed below, the integral of $C_1$ over $S_{x_4}$ at $u\rightarrow\infty$, is fixed to be proportional to $\theta$ by the holographic dictionary. Then, since the $S_{x_4}$ circle shrinks, a trivial solution like $C_1\sim \theta dx^4$ is not allowed. Regularity requires in fact that $C_1$ goes to zero at $u=u_0$, a condition which is precisely satisfied by the solution above. 

The UV 't Hooft coupling and the $\theta$ angle of the gauge theory, can be related to the gravity parameters by considering the low energy limit of the $D4$-brane action 
\be
S_{D4} = - \tau_4 {\rm Tr} \int d^4x\, dx_4 e^{-\phi}\sqrt{-\det (G+ {\cal F})} + \tau_4 \int C_5 + \frac{1}{2}\tau_4\int C_1 \wedge {\rm Tr} {\cal F}\wedge {\cal F}\,,
\ee 
where ${\cal F_{\alpha\beta}}\equiv 2\pi\alpha' F_{\alpha\beta}$ is proportional to the gauge field strength, $\tau_4 = (2\pi)^{-4}l_s^{-5}$, $C_5$ is the electric five-form sourced by the branes (its field strength $F_6$ is the Hodge dual to $F_4$) and $G_{\alpha\beta}$ is the induced metric on the world-volume. 
Expanding the action to second order in derivatives, considering the UV asymptotics $u\rightarrow\infty$, integrating over the compact $x_4$ direction and comparing the resulting 4d action with the Minkowski version of (\ref{lagthetaE}), one gets\footnote{Remember that $\int {\rm Tr} F\wedge F \equiv (1/2)\int d^4x {\rm Tr} F\tilde F$.}
\be
\lambda = g^2 N_c \equiv 2 \lambda_4\equiv 2g_{YM}^2N_c = 4\pi g_s N_c l_s M_{KK}\,,\qquad  \theta + 2\pi k=\frac{1}{l_s} \int_{S_{x_4}} C_1\,,
\label{hd}
\ee
where $k$ is an integer. In the above expression we have introduced the parameters $g_{YM}$ and $\lambda_4$ which are often referred to as the (UV) 4d gauge and 't Hooft coupling of the holographic model in the literature, despite the fact that they differ for a factor of $2$ from the standard bare ones appearing in eq. (\ref{lagthetaE}). As we have observed above, $\lambda_4\sim T_s/M_{KK}^2$ parameterizes how far the holographic model is from pure Yang-Mills. 
 
The second relation in (\ref{hd}) defines $\theta$ mod $2\pi$ integer shifts (since the integral of $C_1$ is gauge invariant only modulo $2\pi{\mathbb Z}$) realizing the expected multi-branched feature of the dual gauge theory vacuum. Moreover, together with (\ref{dilatonu}), it implies that the bare $\theta$ angle is related to the $\Theta$ parameter of the background by
\be
\Theta\equiv\frac{\lambda_4}{4\pi^2}\left(\frac{\theta+2k\pi}{N_c}\right)=\frac{\lambda}{8\pi^2}\left(\frac{\theta+2k\pi}{N_c}\right)\,.
\ee
As it is also suggested by eq. (\ref{lagthetaE}), in the 't Hooft limit, the corrections to the physics due to the $\theta$ parameter, w.r.t. the $\theta=0$ case, are actually weighed by the combination appearing in $\Theta$.  Moreover, since this parameter depends on $k$, what we actually get on the gravity side is an infinite family of solutions corresponding to possible field theory vacua. 

As it was shown in \cite{dubovsky} the curvature invariants of the background remain small if $|\Theta|\ll \lambda_{4}^{1/4}$. A similar constraint is obtained by studying the tension of the domain walls (identified with wrapped $D6$-branes) separating the various $k$-vacua in the dual field theory: if $|\Theta|$ is too large, the (metastable) vacua become unstable. Our results will thus be reliable provided $\Theta$ satisfies these bounds. Actually, as we will see in the following Section, on the field theory ground-state the possible values of $|\Theta|$ turn out to be bound after minimizing the multibranched energy density. 

Let us conclude this Section by recalling that there is an alternative way of getting the holographic relations (\ref{hd}). The one-instanton action $8\pi^2/g^2 + i\theta$ in the gauge theory, is mapped into the Euclidean on-shell action for a $D0$-brane wrapping the $S_{x_4}$ circle.\footnote{This relation is readily obtained by considering the on-shell value of the Euclidean version of the $D4$-brane action written above, on a 4d one-instanton solution $F =\tilde F$ (the solution $F=-\tilde F$ corresponds to an anti-$D0$-brane), with $\int d^4x {\rm Tr}F{\tilde F}=16\pi^2$. Indeed, the $\int C_1 \wedge{\rm Tr} F\wedge F$ term gives one unit of $D0$-brane charge on-shell, since $\tau_0=(2\pi l_s)^4 \tau_4$.} 
In Minkowski, the relevant $D0$-brane action is given by
\be
S_{D0} = -\tau_0 \int dx_4 e^{-\phi} \sqrt{g_{44}}+\tau_0 \int C_1\,,
\ee
where $\tau_0=l_s^{-1}$ is the $D0$-brane charge. 
Using the $u\rightarrow\infty$ limit of the 10d background introduced above, going to the Euclidean frame, and performing the identification with the one-instanton action, one precisely gets the relations (\ref{hd}).
\section{Observables}\label{obs}
\setcounter{equation}{0}
In this Section, using standard holographic methods, we will extract relevant information on the physics of the Yang-Mills theory dual to the $\theta$-backreacted background given above. The main aim is to study the $\theta$ dependence of interesting physical observables, with an eye to the available results for pure Yang-Mills on the lattice (see e.g. \cite{vicari}). As we have already stated in the Introduction, studying the physics at finite $\theta$ is challenging on Euclidean lattices due to the fact that, as shown by eq. (\ref{lagthetaE}), the $\theta$ term in the Euclidean Lagrangian is imaginary. Lattice results are thus obtained either computing the coefficients in series expansions around $\theta=0$ or by extrapolations from imaginary $\theta$ angle, again around $\theta=0$. Correspondingly only the first few terms in powers of $\theta$ are generically computable.\footnote{Effects of the $\theta$ term in certain Yang-Mills theories can be studied with semi-classical methods \cite{Unsal:2012zj}.}

In the holographic model, instead, in the limits where the solution is reliable, the physics can be easily studied and the results exactly given at any order in $\Theta$. Focusing on the small $\theta$ regime where a comparison with lattice results is sensible, we will see that the physics of the holographic Yang-Mills model precisely matches, at least qualitatively, with that of lattice Yang-Mills. This coincidence let us try to adopt the holographic model as a way to predict the behavior of the subleading order terms in the $\theta$ expansion, for the realistic model.  

Moreover, for certain quantities, like the entanglement entropy, which cannot be computed yet on the lattice, not even at $\theta=0$, the results we obtain for the holographic model are unique and could hopefully provide useful benchmarks for pure Yang-Mills.
\subsection{The ground-state energy}
One of the basic entries in the holographic dictionary is the relation between the field theory partition function and the (renormalized) on-shell gravity action. The ground-state energy density $f(\theta)$ of the Yang-Mills theory can thus be obtained through the relation (valid at large $N_c$ and $\lambda_4$)
\be
Z = e^{-V_{4} f(\theta)}\approx e^{-S^{\rm ren}_{E\,{\rm on-shell}}}\,,
\label{diction}
\ee
where $V_4$ is the (infinite) 4d Euclidean spacetime volume and the renormalized on-shell Euclidean gravity action is given by (see also \cite{smearedSS} for a recent account of the holographic renormalization in the model at hand)
\be
S^{\rm ren}_{E\,{\rm on-shell}}= S_{E}+S_{GH}+S_{c.t.}\,.
\ee
Here
\be
S_E=-\frac{1}{2k_0^2}\left[\int d^{10}x\sqrt{g}\left[e^{-2\phi}\left(R+4(\partial\phi)^2\right) -\frac12 |F_4|^2\right]-\frac12 |F_2|^2\right]\,,
\ee
is the Euclidean version of the action (\ref{action1}). Moreover
\be \label{gibbonshawking}
S_{GH} =   -\frac{1}{k_0^2}\int d^{9}x\sqrt{h}e^{-2\phi}K\,,
\ee
is the Gibbons-Hawking term, where $h$ is the determinant of the boundary metric (the slice of the 10d metric (\ref{metricu}) at fixed $u=u_{\Lambda}$ with $u_{\Lambda}\rightarrow\infty$ being the radial position of the UV boundary) and $K$ is the trace of the extrinsic curvature of the boundary
\be
K =  \frac{1}{\sqrt{g}}\partial_u\left(\frac{\sqrt{g}}{\sqrt{g_{uu}}}\right)|_{u = u_{\Lambda}}\,.
\ee
Finally, the counter-term action is given by \cite{myersmateos}
\be \label{bulkcounterterm}
S^{bulk}_{c.t.}=\frac{1}{k_0^2}\left(\frac{g_s^{1/3}}{R}\right)\int d^9x \sqrt{h}\,\frac52 e^{-7\phi/3}\,.
\ee
Evaluating the above terms on the $\theta$-backreacted background given above, one discovers that there are no $\theta$-dependent divergent terms and that the final result reads formally as that for $\theta=0$
\be
S^{\rm ren}_{E\,{\rm on-shell}}= - \frac{1}{2k_0^2g_s^2}V_4\frac{2\pi}{M_{KK}}V_{S^4}\frac{u_0^3}{2}\,.
\label{ren0}
\ee
The $\theta$-dependence comes out explicitly through the relation (\ref{mkk}). Using the holographic map (\ref{diction}) and expressing everything in terms of field theory quantities, from (\ref{ren0}) it follows that the ground-state energy density of the Yang-Mills theory is formally given by (see also \cite{dubovsky}) 
\be
f(\Theta)= -\frac{2 N_c^2 \lambda_4}{3^7 \pi^2}\frac{M^4_{KK}}{(1+\Theta^2)^3}\,,
\label{en0}
\ee
where the negative sign is in agreement with the negative Casimir energy one expects to arise from the compactification of a supersymmetric theory with supersymmetry breaking boundary conditions \cite{HM}.

\begin{figure}[t]
\centering
\includegraphics[width=120mm]{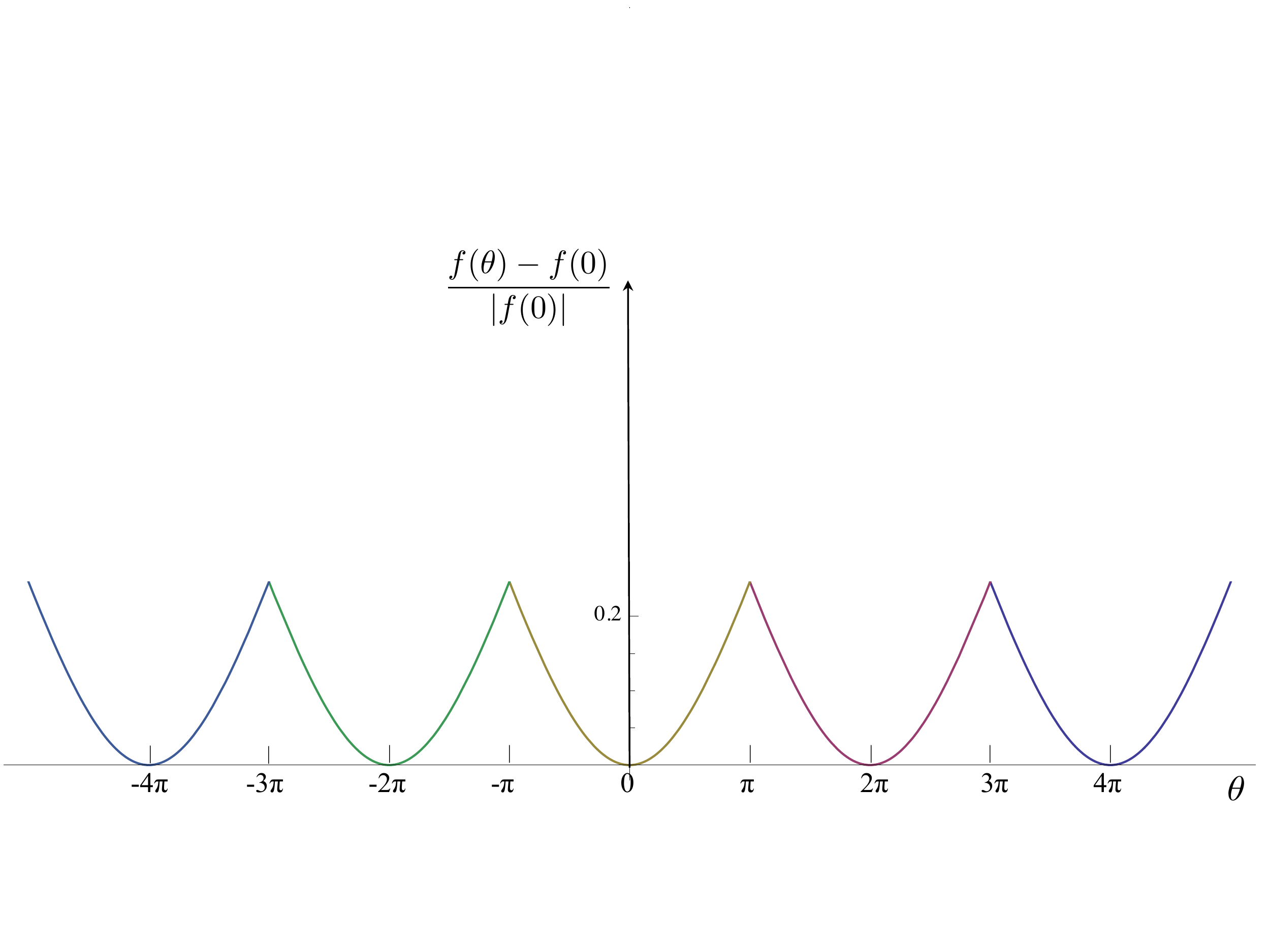}
\caption{\small{The (normalized) ground state energy density at finite $\theta$. Different colors (online) correspond to different branches of the vacuum energy: $k=0$ when $\theta\in(-\pi,\pi)$, $k=1$ when $\theta\in(-3\pi,-\pi)$, $k=-1$ when $\theta\in(\pi,3\pi)$ and so on. At $\theta=\pm\pi$ and odd multiples of these values, a CP breaking first order phase transition separates the different $k$-branches. The plot has been obtained from the function in the text setting $\lambda_4/(4\pi^2 N_c)=0.1$.}}
\label{fig1}
\end{figure}

Actually, since $\Theta$ is proportional to $\theta+2k\pi$, for a given value of $\theta$ the true vacuum energy is obtained by minimizing the previous expression over $k$
\be
f(\theta) = {\rm min}_k f(\Theta)\,.
\ee
As a result, the ground state energy density (see figure \ref{fig1}) turns out to be a periodic function of $\theta$, as expected. To any given interval, of length $2\pi$, of possible values of $\theta$, it will correspond a precise value of $k$. For example, $k=0$ when $\theta\in(-\pi,\pi)$, $k=1$ when $\theta\in(-3\pi,-\pi)$, $k=-1$ when $\theta\in(\pi,3\pi)$ and so on. All in all, in the ground-state $|\theta+2k\pi| < \pi$, so that $|\Theta|<\lambda_4/(4\pi N_c)$.

The theory experiences a first order CP-breaking phase transition when passing from one branch to another. The transitions happen at $\theta=\pm\pi$ and odd multiples of these values: CP symmetry, which would be preserved at these points, breaks spontaneously by the choice of a branch.
In the small $\theta$ limit (which corresponds to the $k=0$ branch) one gets 
\be
f(\theta) - f(0) 
=\frac12\chi_g \theta^2\left[1+{\bar b}_2\frac{\theta^2}{N_c^2}+{\bar b}_4\frac{\theta^4}{N_c^4} + {\cal O}(\theta^6)\right]\,,
\ee
with the topological susceptibility given by \cite{witten}
\be
\chi_g = \frac{\lambda_4^3 M_{KK}^4}{4(3\pi)^6}\,,
\ee
and the expansion coefficients given by
\be
{\bar b}_2 = - \frac{\lambda_4^2}{8\pi^4}\,,\quad {\bar b}_4 = \frac{5\lambda_4^4}{384\pi^8}\,.
\ee
The ${\bar b}_{2n}$ coefficients provide relevant physical information as they are related to the zero-momentum n-point connected correlation functions of the topological charge density at $\theta=0$ \cite{vicari}. These are also phenomenologically interesting quantities: for instance ${\bar b_2}$ (for $N_c=3$) is related to the $\eta'-\eta'$ elastic scattering amplitude. In turn, the ${\bar b}_{2n}$ coefficients give the  moments of the probability distribution of the topological charge $Q$. Non-vanishing coefficients imply that this distribution departs from a simple Gaussian one.

It is interesting to compare these results with those obtained for pure Yang-Mills on the lattice (see \cite{vicari,vicaripos}). Calling $b_{2n}$ (unbarred) the coefficients multiplying just the $\theta^{2n}$ powers, the following values have been found for $N_c=3,4,6$
\be
b_2(\rm lattice) = -0.026(3)\,,-0.013(7)\,,-0.008(4)\,,
\ee
yielding a large $N_c$ estimate for the barred coefficient
\be
{\bar b}_2 \approx -0.2\,.
\ee
It is remarkable to notice that the sign of this coefficient precisely matches with that obtained in the holographic model. 
 
Concerning the value of $b_4$ on the lattice, at present the errors are such that its sign cannot be safely determined: recent results \cite{t41} just provide a bound on its absolute value $|b_4|<0.001$. Provided the qualitative matching of the holographic model and the pure Yang-Mills theory persists at subleading order, our results suggest that $b_4$ should have a positive sign.
\subsection{Rectangular Wilson loop: the string tension}
As it was pointed out in \cite{wilson}, in the large $N_c$, large 't Hooft coupling regime the VEV for a Wilson loop on a contour ${\cal C}$ is holographically given as
\be
\langle W[{\cal C}]\rangle \sim e^{- S^r_{NG}}\,,
\label{Wilsonh}
\ee
where $S^r_{NG}$ is the renormalized Nambu-Goto on-shell action for a fundamental open string whose end-points span the contour ${\cal C}$. For a rectangular contour with sides of length $T$ along the $x^0=t$ direction and length $L$ along one space direction, say $x=x^1$, computing the Wilson loop allows to obtain the static quark-antiquark potential $V(L)$. In the large $L$ limit in fact 
\be
\langle W[{\cal C}]\rangle \approx e^{- T V(L)}\,.
\ee 
Linear confinement of the chromoelectric flux tube implies that $V(L) = T_s L$ at large $L$: the Wilson loop has an area law and $T_s$ is the string tension.

The holographic computation of the rectangular Wilson loop proceeds as follows. Since we are interested in comparing the result with that obtained in pure Yang-Mills on the lattice, we just consider a Wilson loop carrying no spurious global symmetry charges: the corresponding open string will thus be point-like in the transverse $S^4$ as well as in $S_{x_4}$. The relevant string embedding can thus be chosen, in the static gauge, as $\tau=x^0\in[0,T]$, $\sigma=x\in[-L/2,L/2]$, $u=u(x)$ so that the Nambu-Goto action is given by
\be
S_{NG}= - \frac{1}{2\pi\alpha'}\int d\tau d\sigma \sqrt{-g_{\tau\tau}g_{\sigma\sigma}}=-\frac{1}{2\pi\alpha'}T \int dx \sqrt{- g_{00}(g_{xx} + g_{uu}u'(x)^2)}\,,  
\ee
where $g$ is the (induced) string frame metric. The Euler-Lagrange equations for $u(x)$ give the actual string profile to be used for computing the on-shell action. As it has been largely discussed in the literature, for the kind of background we are considering, in the large $L$ limit the string profile is bath-tube-shaped: in order to minimize its energy the string wants to stretch as much as possible over the $u=u_0$ region where its effective tension, proportional to $\sqrt{-g_{00}g_{xx}}$, is minimal. Then, around the extrema $x=-L/2$ and $x=L/2$, the string will move approximately vertically up to the UV cutoff $u=u_{\Lambda}\rightarrow\infty$ where it is attached to a probe brane. The two vertical branches of the string give divergent contributions to the action when the cutoff is sent to infinity. These contributions can be interpreted as due to the (infinite) masses of the static quark-antiquark pair and must be subtracted in order to get the renormalized Nambu-Goto action. All in all, in the large 
$L$ limit one gets
\be
S^r_{NG}\approx -\frac{1}{2\pi\alpha'} T \sqrt{-g_{00}g_{xx}}|_{u=u_0} L\,,
\ee
which, using the holographic map ({\ref{Wilsonh}), implies that the Wilson loop obeys an area law, with a string tension given by
\be
T_s = \frac{1}{2\pi\alpha'}\sqrt{-g_{00}g_{xx}}|_{u=u_0}\,.
\ee
On the $\theta$-deformed background (\ref{metricu}) this reads\footnote{In \cite{Bigazzi:2004ze} the leading stringy corrections to this result have been derived.  Following the rule for the $\theta$ corrections mentioned in the Introduction, the full results would then read \be
T_s = \frac{2\lambda_4}{27\pi}M_{KK}^2\frac{1}{(1+\Theta^2)^2}\left(1- \frac{27}{4\lambda_4}(1+\Theta^2)\log{2} \right)\,.
\ee}
\be
T_s = \frac{2\lambda_4}{27\pi}M_{KK}^2\frac{1}{(1+\Theta^2)^2}\,.
\label{tsh}
\ee
As it has been shown in the previous Section, one has to read this expression carefully: actually $\Theta=\theta+2k\pi$ and, in the ground-state, for a given value of $\theta$ in an interval of length $2\pi$ one has to fix the corresponding value of $k$.

Setting $k=0$ and considering the $\theta\rightarrow0$ limit one gets
\be
T_s =\frac{2\lambda_4}{27\pi}M_{KK}^2\left(1 - \frac{\lambda_4^2}{8\pi^4}\frac{\theta^2}{N_c^2} +\frac{3\lambda_4^4}{256\pi^8}\frac{\theta^4}{N_c^4}+ {\cal O}(\theta^6)\right)\,.
\label{ts}
\ee
It is remarkable that the ${\cal O}(\theta^2)$ correction is negative. This, together with the expected scaling with $\theta/N_c$, is precisely what has been found on the lattice in \cite{vicari}
\be
T_{s\,lat} = T_{s\,lat}(0)\left[1+{\bar s}_2\frac{\theta^2}{N_c^2}+\dots\right]\,,
\ee
where, using the large $N_c$ extrapolation of the $N_c=3,...,6$ results one gets\footnote{Calling $s_2$ the coefficient multiplying just $\theta^2$, on the lattice on finds, for $N_c=3$,  $s_2 = -0.08(1)$.}
\be
{\bar s}_2\approx -0.9\,.
\ee 
Moreover, using the holographic result (\ref{ts}) we are led to predict that the ${\cal O}(\theta^4)$ correction to the string tension in pure Yang-Mills should be positive.
\subsection{'t Hooft loop and oblique confinement}
The rectangular 't Hooft loop computes the monopole-anti-monopole potential. At $\theta=0$ one expects that confinement of the chromoelectric flux is associated with screening of the magnetic charges. The situation is expected to change, instead, when the $\theta$ angle is turned on. Let us see what the holographic results tell us about this.

The ``chromomagnetic string" in the model we are considering is a probe $D2$-brane wrapped on $S_{x_4}$ \cite{brandh, Gross:1998gk}. In order to follow standard conventions, it is useful to write the relevant part of its action as
\be
S_{D2} = - T_2 \int d^3\xi e^{-\hat\phi}\sqrt{-\det (g + {\cal F})} + T_2 \int {\hat C}_1\wedge {\cal F}\,,
\ee
where $T_2=g_s^{-1}\tau_2=(2\pi)^{-2}g_s^{-1}l_s^{-3}$ is the $D2$-brane tension, $e^{-\hat\phi}\equiv g_s e^{-\phi}$, ${\hat C}_1\equiv g_s C_1$ and ${\cal F}=2\pi\alpha' F$ is a $U(1)$ gauge field on the brane world-volume. Notice that due to the non-trivial $C_1$ potential in the background, the equations of motion for $F$ are not solved by the trivial solution $F=0$. Instead of working out these equations, it is easier to make use of the relation between the action above and that of a wrapped $M2$-brane in the 11d completion of the background (as we will see, only the low energy details of the latter play a role in the computations, hence working with the M-theory completion is equivalent to working with the 10d one). This relation can be obtained using the following procedure (see e.g. \cite{m2d2}). 

One introduces an auxiliary vector field $t_{\alpha}$ on the $D2$-brane world-volume and rewrites the action as
\be
{\tilde S}_{D2} = -T_2 \int d^3\xi\left[e^{-\hat\phi}\sqrt{-\det (g_{\alpha\beta} + e^{2\hat\phi}t_{\alpha}t_{\beta})} - \frac12\epsilon^{\alpha\beta\gamma}({\hat C}_{\alpha}- t_{\alpha}){\cal F}_{\beta\gamma}\right]\,.
\ee
Integrating out $t_a$ one gets back the original action. If instead we treat the gauge field ${\cal A}$ in ${\cal F}=d{\cal A}$ as a Lagrange multiplier we see, from its equation of motion, that 
\be
\epsilon^{\alpha\beta\gamma}\partial_{\beta}({\hat C}_{\gamma}-t_{\gamma})=0\,,
\ee
which implies that ${\hat C}_{\alpha}-t_{\alpha}=\partial_{\alpha} y$ with $y$ being a scalar. Using this equation one discovers that the $D2$-brane action becomes equivalent to the $M2$-brane action
\be
S_{M2} = -\frac{1}{(2\pi)^2 l_{11}^3} \int d^3\xi\sqrt{-\det G}\,,
\label{m2a}
\ee
where $l_{11}\equiv l_s g_s^{1/3}$ and $G$ is the pullback on the brane world-volume of the 11d metric
\be
ds_{11}^2 = e^{-\frac23\hat\phi}ds_{10}^2 + e^{\frac43\hat\phi}(dy- {\hat C}_{x_4} dx_4)^2\,.
\label{11dme}
\ee
Here $ds_{10}^2$ is the 10d string frame metric given by (\ref{metricu}) and $y$ is identified with the compact eleventh direction of M-theory of length $2\pi R_{y}\equiv 2\pi g_s l_s$, compactifying over which one gets back the 10d IIA string theory background.
In 11d the whole background reduces to the metric above and a four form field strength, which solve the equations of motion of the 11d supergravity action. 

As already pointed out in \cite{Gross:1998gk}, there are two relevant cycles one can identify in the $(x_4,y)$ subspace. The first one is defined by the equation $x_4=const$: this is the cycle over which one reduces M-theory to get the type IIA model. An $M2$-brane wrapped over this cycle reduces to a fundamental string. The second cycle is what we are interested in: it is the minimal volume contractible cycle over which we have to wrap the $M2$-brane to get the chromomagnetic string. Looking at the UV ($u\rightarrow\infty$)  asymptotics of the background, we see that this cycle is defined by 
\be
y=\Theta x_4 + const\,.
\ee
The metric on this cycle reads, thus
\be
ds^2 = \left[e^{-\frac23\hat\phi}g_{44}+ e^{\frac43\hat\phi}\Theta^2\left(1-\frac{f}{H_0}\right)^2\right] dx_4^2\,.
\ee
Notice that the volume of this cycle is minimized at $u=u_0$, where it takes a {\it finite} value if $\Theta$ is different from zero. In this case, thus, the wrapped $M2$-brane action, in the limit where the monopole-antimonopole distance $L$ is very large, is minimized by a bath-tube shaped configuration, precisely as in the case of the Wilson loop fundamental string. Using the same logic as in that case, one gets, for the chromomagnetic string tension, the following remarkable expression
\be
T_m = \frac{1}{27\pi^2} M_{KK}^2 \lambda_4 \frac{|\theta+2k\pi|}{(1+\Theta^2)^2} \equiv T_s \frac{|\theta+2k\pi|}{2\pi}\,.
\ee
Hence, at any order in $\Theta$ within our approximations, there is a very simple relation between the chromomagnetic tension $T_m$ and the string tension $T_s$. 
Notice that at $\theta=0$ ($k=0$) this relation implies that $T_m=0$: as expected there is no area law for the 't Hooft loop and the magnetic monopoles are screened. 
At finite $\theta$, instead, the 't Hooft loop generically shows an area law. The objects which are screened are actually dyons, particles of electric charge $-p$ and magnetic charge $q$. 
Using the above relation we see that the ``string'' tension for the dyons is (on the $k=0$ branch and taking $\theta>0$)
\be
T_{dy} = -p T_s + q T_m = \left(-p + \frac{\theta}{2\pi}q\right) T_s\,,
\ee
so that dyons are screened provided that $\theta= 2\pi (p/q)$: this is precisely the so-called oblique confinement. 
\subsection{The scalar glueball mass}
Confinement in Witten's holographic model is accompanied, just as in pure Yang-Mills, by the occurrence of a mass gap in the (glueball) spectrum. In pure Yang-Mills the lightest glueball is a $C$ and $P$ even scalar mode $0^{++}$. 
The $\theta$ dependence of its mass has been studied on the lattice in \cite{Del Debbio:2006df}: there, the $0^{++}$ mass was found to decrease quadratically with $\theta$ around $\theta=0$. 
As customary, the regime investigated by holographic means is different from the lattice one.
Nevertheless, we are after the qualitative behavior of the observables.
This is hoped to be a stable feature of the theory.
So, let us compare the lattice result with the holographic model.

At $\theta=0$ the spectrum of glueballs in Witten's model has been studied in many papers (for example in \cite{Csaki:1998qr,Constable:1999gb,Brower:2000rp,Elander:2013jqa}).
There are actually two light $0^{++}$ modes.
The lightest one was dubbed ``exotic'' in \cite{Constable:1999gb}, as it comes from a metric perturbation involving, among many others, the compact $x_4$ direction of the background; it is sometimes called ``mode S'' in the literature.
The second light mode was dubbed ``dilatonic'' in \cite{rebhan}, as it involves a fluctuation of the dilaton in the ten dimensional geometry; it is customarily called ``mode T''.
It is degenerate in mass with the tensorial $2^{++}$ and vectorial $1^{++}$ glueballs, due to the symmetries of the background - it comes from the extra flat Minkowski direction in the eleven dimensional origin of the background, so the equation governing its mass is precisely the same as the one for the Minkowskian tensorial perturbations.

Since it is lighter, not degenerate with the $2^{++}$ and common to some deformations of Witten's model \cite{Elander:2013jqa}, at first sight it would seem that the ``exotic'' mode S is to be considered in the comparison with Yang-Mills data.
This could actually not be the case.
As a first point, from inspection of the DBI action of a $D4$-brane, one can see that the operator $Tr F^2$ sources both the S and T $0^{++}$ modes.
Furthermore, the ``exotic'' polarization could not survive in the limit where one decouples the KK modes, as it comes primarily from the excitations along the KK direction $x_4$.
Moreover, the extrapolation of lattice data at large $N_c$ is still unprecise \cite{Elander:2013jqa}.
Finally, a recent study of glueball decays has shown that the ``exotic'' mode is too broad (and too light) to be compatible with the scalar glueball candidates in QCD \cite{rebhan}.\footnote{See also \cite{Brunner:2015yha}.}
All in all, although the situation is not firmly settled yet, it seems possible, if not likely, that it is the ``dilatonic'' T mode and not the ``exotic'' S mode the correct one to be compared with QCD - it should be the only one sourced by $Tr F^2$ in the KK mode decoupling limit.
For these reasons, we will concentrate on the ``dilatonic'' T mode in the following.

It is convenient to work in eleven dimensions, so that the only field involved in the discussion is the metric. The $\theta$-backreacted background it is given by eq. (\ref{11dme}). The general equation for the metric fluctuations (also at finite $\theta$), obtained by linearizing the 11d Einstein equations on the background, reads \cite{Constable:1999gb}
\be\label{fluctuations}
\frac12 \nabla_a \nabla_b h_c^c + \frac12 \nabla^2 h_{ab} - \nabla^c \nabla_{(a} h_{b)c} -\frac64 h_{ab}=0\,.
\ee
Here $a, b, c, ...$ are seven dimensional indexes referring to the reduction of the eleven dimensional background (\ref{11dme}) on the four sphere, while $\mu, \nu, ...=0, ..., 3$ refer to the four Minkowski directions.
The fluctuations can be factorized as $h_{ab}=H_{ab}(u) e^{-ik\cdot x}$, with $k^2=-M^2$; we work in the frame where $k^{\mu}=\omega \delta^{\mu}_t$.

The ``dilatonic'' mode T at $\theta=0$ corresponds to a traceless excitation $H_{ab}=\epsilon_{ab}H(u)$ with diagonal components only along the spatial directions $x_{\mu}$ with $\mu=1,2,3$ and $y$.\footnote{At $\theta=0$ the equation of motion for this fluctuation coincides with that of a minimally coupled scalar.
}
The key point in our discussion is that it is sufficient to keep the traceless condition also at finite $\theta$ in order for the mode to be compatible with equation (\ref{fluctuations}), provided one single equation for $H(u)$, giving the mass spectrum of the glueballs, is satisfied.
The correct ansatz for the fluctuation is
\be\label{ansatzfluct}
H_{ab}(u)= \frac{u}{R} H(u)\, {\rm diag}\left(0,1,1,1,0,- \frac{3}{1+\Theta^2},0\right)\, , 
\ee
giving the equation
\be\label{eqfluct}
 H''(u) + \frac{4u^3-u_0^3}{u(u^3-u_0^3)} H'(u) - \frac{M^2 R^3}{u^3-u_0^3} H(u)= 0\, .
\ee
This has acceptable solutions (regular at $u=u_0$ and normalizable at $u\rightarrow\infty$) only if $M^2>0$ and the spectrum is discrete. This is the holographic realization of the mass gap \cite{witten}.

The fact that the ansatz (\ref{ansatzfluct}) is sufficient was not a-priori guaranteed, since the metric at finite $\theta$ has non-diagonal entries which are expected to cause a mixing among the $\theta=0$ modes.
Indeed, the fact that generically the mixing happens was found in \cite{gaba}.
There, the backreaction of the RR one-form on the metric was not considered, but taking into account its $\theta$ dependent background value it was shown that the gravity mode dual to the $0^{-+}$ glueball mixes with the T and S modes (while the latter obviously satisfy the $\theta=0$ equation, as the metric was unchanged).
The mixing between the $0^{++}$ and $0^{-+}$ glueballs has to be expected, as $\theta$ breaks CP invariance.
In fact, somewhat surprisingly, in \cite{gaba} it was found that, despite the mixing of the modes, the masses of these glueballs are not changed if the backreaction of $F_2$ on the geometry is not taken into account.

What we have found here is that, despite the full backreaction of $\theta$ on the geometry, the small modification of the ansatz for the mode T in (\ref{ansatzfluct}) still preserves its equation: in fact (\ref{eqfluct}) is identical to the equation at $\theta=0$.
This fact was observed in \cite{dubovsky} for the lightest $2^{++}$ mode (although in that case no mixing is actually expected).
Note that the $2^{++}$ and $0^{++}$ T modes are still degenerate in mass at finite $\theta$.

The fact that the equation for a mode is unchanged by $\theta$ does not imply that the mass of the glueball does not depend on $\theta$.
In fact, the masses are given in units of $u_0/R^2$, whose relation with the physical scale $M_{KK}$ is $\theta$ dependent: from equation (\ref{mkk}) it follows immediately that the $\theta$ correction to the mass of the lightest relevant glueball is given by 
\be\label{emme}
M(\Theta)= \frac{M(\Theta=0)}{\sqrt{1+\Theta^2}}\,. 
\ee
As usual this expression implies that $M(\theta)$ is a periodic function.
Focusing on the $k=0$ branch and expanding around $\theta=0$ one thus finds
\be
M(\theta)= M(\theta=0)\left(1-\frac{\lambda_4^2}{32\pi^4}\frac{\theta^2}{N_c^2}+\frac{3\lambda_4^4}{2048\pi^8}\frac{\theta^4}{N_c^4}+ {\cal O}(\theta^6)\right)\,.
\ee
The leading correction in $\theta^2$ has a negative sign, precisely as it was found in lattice Yang-Mills \cite{Del Debbio:2006df}, where, for $N_c=3$
\be
M_{lat}(\theta) = M_{lat}(0)\left(1+ g_2 \theta^2 + {\cal O}(\theta^4)\right)\,,\quad g_2=-0.06(2)\,.
\ee
The holographic model predicts that the ${\cal O}(\theta^4)$ correction has a positive sign. Again, it would be nice to check this expectation on the lattice.
\subsection{The mass of the baryon vertex}
In Witten's holographic model, a baryon vertex is identified with a $D4$-brane wrapped on $S^4$ and localized at the radial position corresponding to the deep IR of the dual field theory  \cite{Witten:1998xy}. 
Using the $\theta$-backreacted background, we can easily study how a finite $\theta$ term affects the mass of the baryon vertex. The wrapped $D4$-brane action reads
\be\label{baryonvertex}
S_{D4}= -\tau_4 \int dx^0 d\Omega_4 e^{-\phi}\sqrt{\det g_5}|_{u=u_0}\equiv -m_B\int dx^0\,,
\ee 
where $\tau_4=(2\pi)^{-4}l_s^{-5}$ is the $D4$-brane charge.
Using formulas (\ref{metricu})-(\ref{f42}) in (\ref{baryonvertex}), integrating on $S^4$ and substituting the expressions in (\ref{mkk}), (\ref{hd}), one finds that the baryon vertex mass is given by
\be
m_B(\Theta) =  \frac{\lambda_4}{27\pi}N_c M_{KK}\frac{1}{(1+\Theta^2)^{3/2}}\equiv \frac{m_B(0)}{(1+\Theta^2)^{3/2}}\,,
\ee
which is a periodic function of $\theta$. Thus, the mass of the baryon vertex decreases with $\theta$.
\subsection{Finite temperature: the confinement-deconfinement transition}
Going to finite temperature in Witten's holographic Yang-Mills model, amounts to compactifying the Euclidean time on a circle of length $\beta=1/T$. As in the $\theta=0$ case, there are two possible solutions of the gravity equations of motion for which this condition can be satisfied (see e.g. \cite{Aharony:2006da}).

One solution has precisely the same structure of the one discussed above, the only difference being just in the fact that its Euclidean continuation has a compact time circle. This solution corresponds to the confined phase of the theory. The free energy density (hence minus the pressure) in this phase has the same expression as in the $T=0$ case (\ref{en0}) 
\be
 f = -p = - \frac{2 N_c^2 \lambda_4}{3^7 \pi^2}\frac{M^4_{KK}}{(1+\Theta^2)^3}\equiv \frac{f(0)}{(1+\Theta^2)^3}\,.
\label{pressure}
\ee
The usual gravity solution corresponding to the deconfined phase has a black hole event horizon and the $x_4$ circle does not shrink anymore.\footnote{In \cite{Mandal:2011ws} it has been proposed an alternative solution for the deconfined phase and \cite{Hanada:2015gsa} presents a study of instantons on that background.}
This allows $F_2=0$, i.e. $C_1 = \theta dx^4$ to be a solution of the background equations of motion. As a result the background metric does not depend on $\theta$ and it is given by
\bea\label{wittenT}
&& ds^2 = \left(\frac{u}{R}\right)^{3/2}\left[{\tilde f}(u) dx_0^2 + dx_a dx^a + dx_4^2\right] + \left(\frac{u}{R}\right)^{-3/2}\left[\frac{du^2}{{\tilde f}(u)} + u^2 d\Omega_4^2\right]\,,\nonumber \\
&& {\tilde f}(u) = 1- \frac{u_T^3}{u^3}\,.
\eea
The dilaton and the $F_4$ form do not change w.r.t. the $T=0$ solution. In the equation above, $a=1,2,3$. The $(x_0, u)$ subspace is topologically a disk - with the $S_{x_0}$ circle smoothly shrinking to zero size at $u=u_T$ (the position of the horizon) - provided we identify
\be
9 \beta^2 u_T = 16\pi^2 R^3\,.
\label{temp0}
\ee
The fact that this background actually corresponds to a deconfined phase of the dual field theory is readily seen recalling, for example, what we got for the Wilson loop: the would-be string tension in this case would be zero since $g_{00}g_{xx}|_{u=u_T}=0$. 

Using the Bekenstein-Hawking formula one gets the black hole entropy density (which is holographically mapped to that of the dual field theory)
\be
s = \frac{256 N_c^2 \pi^4 \lambda_4}{729 M_{KK}^2}T^5\,,
\ee
from which we can deduce the free energy density (hence minus the pressure) using $s=-(\partial f/\partial T)$ 
\be\label{freeen}
f _{dec}= - p_{dec} = - \frac16 \frac{256 N_c^2 \pi^4 \lambda_4}{729 M_{KK}^2} T^6\,.
\ee
Notice the peculiar scaling with $T$, which follows from the higher dimensional UV completion of the model. Since the free energy is independent from $\theta$, one gets that in the holographic model at hand, the topological susceptibility is zero in the deconfined phase. This is actually what one expects in the large $N_c$ limit of Yang-Mills, where the susceptibility is exponentially suppressed.\footnote{In the deconfined high temperature phase the topological sector of the Yang-Mills theory is well described in terms of an instanton gas. The same description does not hold anymore in the confined phase. This is encoded in the dual holographic picture (see e.g. \cite{Bergman:2006xn,pz}
 and references therein). As we have seen, instantons are identified with $D0$-branes wrapping the $S_{x_4}$ cycle. In the deconfined phase, this cycle does not shrink and the $D0$-brane configuration is stable. In the confined phase, instead, the $S_{x_4}$ circle shrinks and the same configuration is unstable.} 
 
Comparing the free energies (or pressures) in the two allowed phases on gets that the dual field theory experiences a first order confinement-deconfinement phase transition at a critical temperature $T_c$ obtained by solving the equation $p=p_{dec}(T_c)$ where the pressure $p$ in the confined phase is the one given in (\ref{pressure}). The result is
\be
T_c(\Theta) = \frac{M_{KK}}{2\pi}\frac{1}{\sqrt{1+\Theta^2}}\,.
\label{tc}
\ee
As usual we have to recall that $\Theta$ is actually proportional to $\theta+2k \pi$ and that, as we have observed studying the vacuum energy density, a given interval (of length $2\pi$) of possible values of $\theta$ is related to a precise value of $k$. 
As a result, plotting the actual critical temperature $T_c(\theta)$ one gets a periodic behavior like that shown in figure \ref{fig2}. 
It is remarkable that the holographic model explicitly realizes the expected but still conjectural (see e.g. \cite{massimo}) ``arcade" structure, allowing, in turn, to study the behavior of the phase diagram near the critical points (which turn out to be triple points) where the first order deconfinement transition line meets the first order lines related to the CP breaking transitions in the confined phase.\footnote{Let us focus for example on that at $\theta=\pi$. It is easy to show that the phase diagram at that point has a cusp: when $\theta\rightarrow\pi^{\pm}$, $T_c(\theta)/T_c(0)\approx (T_c(\pi)/T_c(0)) + A(\pi\pm\theta)$, where $A>0$ is given in terms of the parameters of the model.} It is tempting to conjecture that the same cusped structure should appear in pure Yang-Mills at large $N_c$.  

Focusing on the $k=0$ branch and expanding around $\theta=0$ we find (at ${\cal O}(\theta^2)$ the result was already found in \cite{smearedSS})
\be
T_c(\theta) = \frac{M_{KK}}{2\pi}\left[1- \frac{\lambda_4^2}{32 \pi^4}\frac{\theta^2}{N_c^2}+\frac{3\lambda_4^4}{2048\pi^8}\frac{\theta^4}{N_c^4}+{\cal O}(\theta^6) \right] \,,
\ee
where, again, both the quadratic dependence and the sign of the leading $\theta$-dependent correction, agree with what has been found on the lattice \cite{massimo}. There, for $N_c=3$ it has been found that
\be
{T_c(\theta)}_{lat} = {T_c(0)}_{lat}\left[1 - R_{\theta}\theta^2 + {\cal O}(\theta^4)\right]\,,\quad R_{\theta}= 0.0175(7)\,.
\ee 
Our model, then, predicts that the ${\cal O}(\theta^4)$ correction has a positive coefficient. It would be nice to check this prediction on the lattice.
\begin{figure}[t]
\centering
\includegraphics[width=100mm]{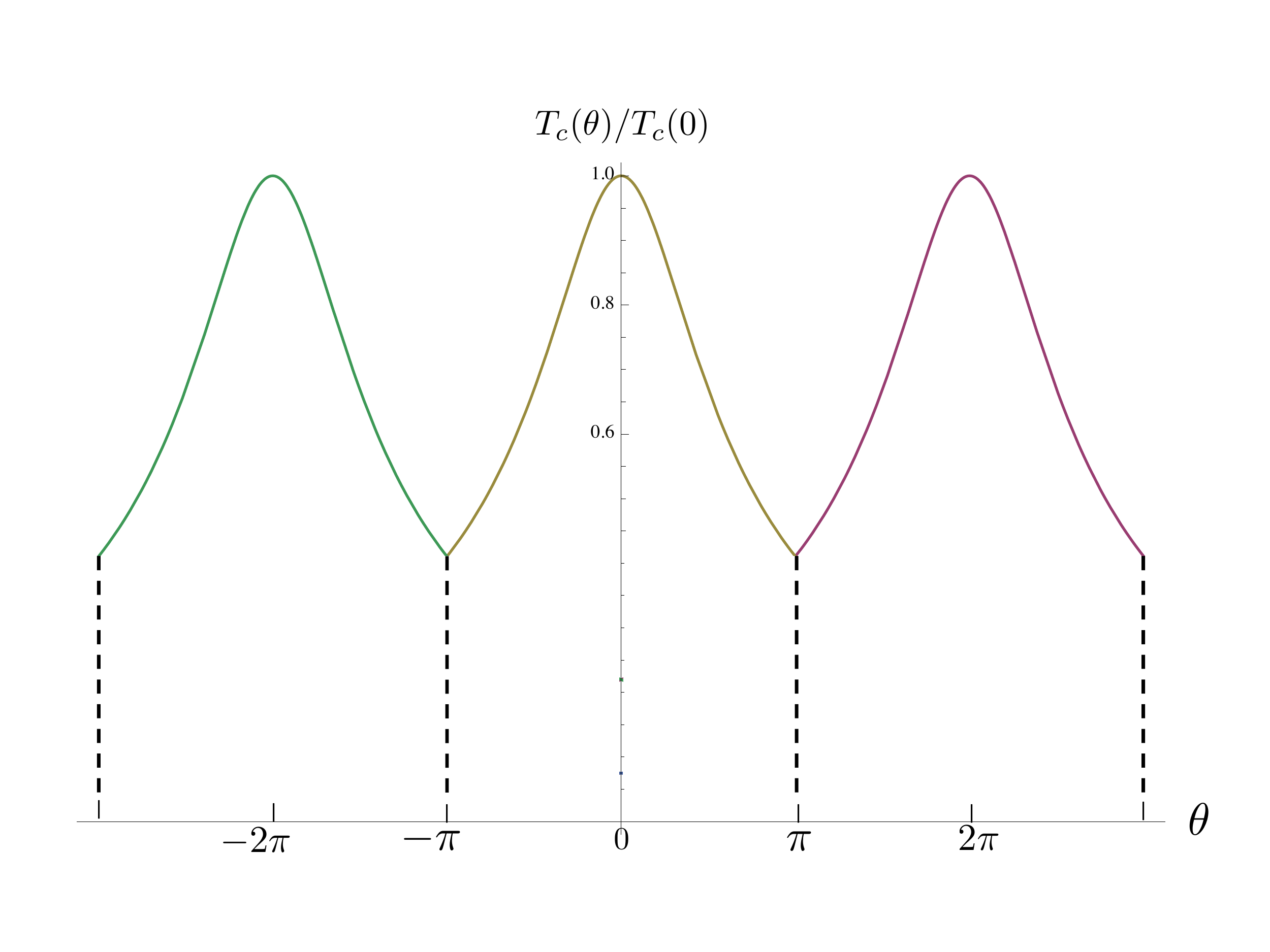}
\caption{\small{The $(T,\theta)$ phase diagram. Different colors (online) correspond to different branches of the vacuum energy: $k=0$ when $\theta\in(-\pi,\pi)$, $k=1$ when $\theta\in(-3\pi,-\pi)$, $k=-1$ when $\theta\in(\pi,3\pi)$ and so on. The critical temperature has cusps at $\theta=\pm\pi$ and odd multiples of these values. The cusps are actually triple points where the deconfinement first order transition line meets the CP breaking first order transition dashed lines separating the different $k$-branches in the confined phase.}}
\label{fig2}
\end{figure}
\subsection{Entanglement entropy}
In this section we compute the dependence of the entanglement entropy on the $\theta$ angle, as a tool to characterize the phases of the theory.
As customary, we take the entanglement entropy as a measure of entanglement between two physically disjoint spatial ($t={\rm const.}$) regions in the boundary theory: region A and its complement, region B.
These regions are separated by a given surface $\gamma$, whose shape can enter the result for the entanglement entropy.
We will mostly consider the ``slab'' and the ball geometries for region A. 
The results will be that in the UV the entanglement entropy is unaffected by the $\theta$ angle, while it is suppressed in the IR; the critical length at which this confining model exhibits a first order transition in the entanglement entropy increases with $\theta$.

The holographic prescription for the entanglement entropy \cite{rt} (see also \cite{rt2,rt3}) consists in calculating the (Einstein frame) area of the minimal (eight dimensional in our setup) bulk surface $\Gamma$ which is anchored to the given surface $\gamma$ at the boundary separating regions A and B. 
This is achieved by minimizing the action (which we write in terms of the string frame metric $g_{MN}=e^{\phi/2}g^{(E)}_{MN}$, since this is the one we have worked with in the previous Sections)
\be
S=\frac{1}{4G_N}\int d^8\sigma e^{-2\phi} \sqrt{-\det(G_{{\rm ind}})} \, , 
\ee
where $G_N= 8 \pi^6 g_s^2 l_s^8$ is the ten dimensional Newton constant and $G_{{\rm ind}}$ is the induced string frame metric on the eight dimensional (fixed time) bulk surface $\Gamma$.

In our background it is straightforward to realize that the form of the metric and dilaton (\ref{metricu}), (\ref{dilatonu}) are such that the factors of $H_0$, carrying the explicit dependence on $\Theta$, cancel out from the expression $e^{-2\phi} \sqrt{-\det(G_{{\rm ind}})}$ for any surface $\Gamma$.

This can be seen also by performing a dimensional reduction (over the compact transverse spaces $S^4$ and $S_{x_4}$) to five dimensions, where the metric does not depend on $H_0$ anymore. The canonically normalized Einstein frame 5d metric reads, in fact
\be
ds^2_{5\,E}=f(u)^{1/3} \left(\frac{u}{R}\right)^{5/3}\left[dx_{\mu}dx^{\mu} + \left(\frac{R}{u}\right)^{3}\frac{du^2}{f(u)}\right]\,,
\label{5dred}
\ee
with the 5d Newton constant defined as $G^{-1}_{N(5)}=G^{-1}_NR^4V_{S^4}2\pi M_{KK}^{-1}$. The holographic entanglement entropy can be thus equivalently deduced by minimizing the area
\be
S_{(5)} = \frac{1}{4 G_{N(5)}}\int d^3\xi \sqrt{-\det G_{(5){\rm ind}}}\,,
\ee
where $G_{(5){\rm ind}}$ is the pullback of the 5d Einstein frame metric (\ref{5dred}) on the surface.

All in all, the dependence on $\Theta$ just follows from the explicit expression of the stringy variables in terms of field theory ones, formula (\ref{mkk}).

As explicit examples we consider the ``slab'' geometry, for which the entanglement entropy has been derived in \cite{klebanov}, and the ball geometry considered in \cite{wen}.
We can just repeat the analysis in those papers, to which we refer the reader for details, with minimal modifications.\footnote{Note that the $\lambda=g_sN_c$ in \cite{klebanov,wen} is related to our $\lambda_4$ by an inverse $2\pi M_{KK}$ factor.}

\subsubsection{The ``slab'' geometry}
This is the simplest geometry to consider: region A is the direct product of $\mathbb R^2$ times an interval of length $l$ \cite{klebanov}.
The minimal surface in the bulk with these boundary conditions has distinct features at small and large $l$.
For small $l$ it is determined by a connected surface extending in the bulk up to a minimal radial position $u_*>u_0$.
For large $l$ the minimal surface is given by two disconnected pieces, anchored at the boundary to the two extrema of the segment and extending in the bulk all the way up to $u_0$.
The critical length $l_c$ at which the (first order) transition between these two configurations happens could be viewed as a probe of the scale of confinement, since the connected surface explores a UV (``deconfined'') region, while the connected one a IR  (``confined'') region.

The entanglement entropy corresponding to the disconnected solution, dominating at large $l$, can be derived analytically and reads
\be\label{eeslabdisc}
S_{dis} = \frac{V_2}{4 G_{N(5)} R}(u_{\infty}^2-u_0^2) = \frac{V_2}{4 G_N} R^3 V_{S^4}\frac{2\pi}{M_{KK}} (u_{\infty}^2 - u_0^2)\,, 
\ee
where $V_2$ is the (infinite) two dimensional volume of $\mathbb R^2$ and $u_{\infty}$ is a UV cut-off.
In terms of field theory variables it can be written in the following form
\be\label{eeslabdiscb}
S_{dis} = V_2 \frac{2^2}{\pi 3^5}\frac{\lambda_4 N_c^2}{(1+\Theta^2)^2}M_{KK}^2 \left(\frac{1}{\epsilon} - 1\right)\,, 
\ee
where $\epsilon=u_0/u_{\infty}\rightarrow0$ accounts for the UV divergence.
$S_{dis}$ has the typical parametric dependence of a 4d theory, which is the case in the IR of the Witten model.
Also, the result does not depend on $l$, so that $\partial S_{dis}/\partial l \sim {\cal O}(N_c^0)$, while it is very sensitive to the $\theta$ angle, which affects the IR of the theory. Notice that, as usual, the previous expression has to be read by taking into account the $\theta, k$ dependence of $\Theta$. As a result, $S_{dis}(\theta)$ is a periodic function.

The connected surface allows to get an analytic form only for small $l$; the UV divergent part of the entanglement entropy is precisely as in (\ref{eeslabdisc}), while its finite part reads
\be\label{eeslabconn}
S_{conn,fin} = -V_2 \frac{2^6}{3}\frac{\pi^{3/2} \Gamma[\frac35]^5}{\Gamma[\frac{1}{10}]^5}\frac{\lambda_4 N^2_c}{(l\; M_{KK})^4} M_{KK}^2\,.
\ee
The dependence on $l$ is characteristic of a 6d theory, consistently with the UV of Witten's theory.\footnote{In \cite{klebanov} it is pointed out that using the fact that $g_s l_s$ is the size $R_{y}$ of the M-theory circle, one can rewrite $\lambda_4 N_c^2/M_{KK} \sim R_{y}N_c^3$ giving the correct behavior of a 6d theory compactified on a circle of length $2\pi R_{y}$.} 
Note that in this case $\partial S_{conn}/\partial l \sim {\cal O}(N_c^2)$ \cite{klebanov}.
Moreover, there is no dependence on $\theta$, consistently with the fact that the UV of the theory is not sensitive to the shift in the vacuum energy.

This fact provides immediately the behavior of the critical length $l_c$ with $\Theta$.
In fact, comparing the two types of solutions, eqs. (\ref{eeslabdiscb}) and (\ref{eeslabconn}), it follows immediately that
\be \label{eecritl}
l_c=l_{c,0} \sqrt{1+\Theta^2}\,, 
\ee
being $l_{c,0}\sim 1.288/M_{KK}$ the critical length at $\Theta=0$. 
This expression is confirmed by the numeric comparison of the entanglement entropy of the disconnected solution with the connected one ($l$ is not forced to be small in this case), see also figure \ref{figE} (showing the swallow-tail trend typical of first order transitions\footnote{A similar behavior, for another non-local observable as the rectangular Wilson loop, has been found in QCD-like holographic models with massive flavors, see e.g. \cite{WLnoi}.}).
\begin{figure}[t]
\centering
\includegraphics[width=100mm]{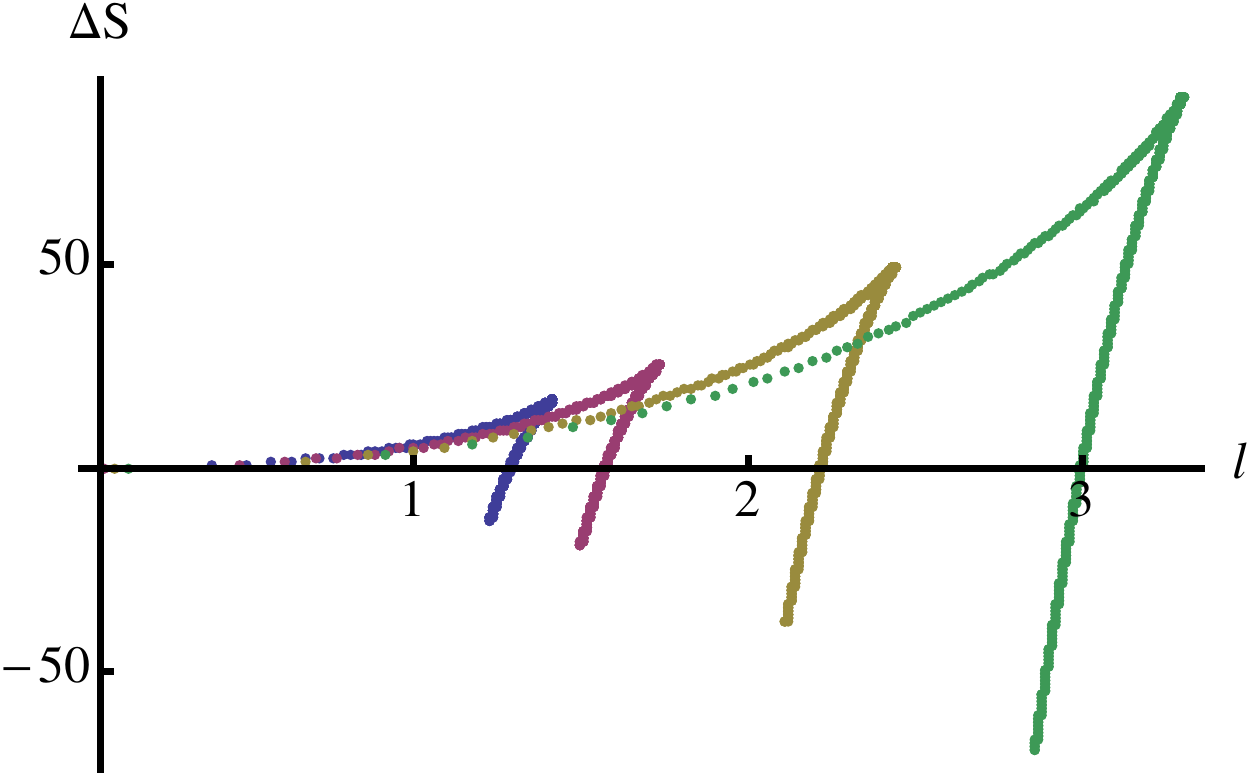}
\caption{\small{The difference between the entanglement entropies corresponding to the connected and disconnected solutions as a function of the length $l$, for $\Theta=0, 0.7, 1.4, 2.1$ (left to right). Notice the typical swallow-tail behavior.}}
\label{figE}
\end{figure}
So, the critical scale for the phase transition, which is a probe of confinement, behaves as $l_c M_{KK}\sim \sqrt{1+\Theta^2}$. 
The usual considerations on the $k,\theta$ correlated values, imply that the actual critical length $l_c(\theta)$ is a periodic function. 

The main effect of the $\theta$ angle is to enhance the vacuum energy (in the ``confined'' phase).
As such, the length at which the UV (``deconfined'') regime is probed is (in each branch) larger at larger $|\theta|$, as signaled by formula (\ref{eecritl}).

\subsubsection{The ball geometry}
In $(2+1)$d the entanglement entropy of a large disc with its complement has a part independent of the disc radius $r$, which provides the so-called ``topological entropy'', a measure of the topological order of a gapped theory.
The natural generalization to $(3+1)$d theories of this quantity, which in this case has no topological nature, is encoded in the entanglement entropy of a ball (with a two-sphere boundary) with its complement.
This has been investigated for the Witten theory in \cite{wen}.
We can follow the latter reference to obtain the dependence of the entanglement entropy on the $\theta$ angle.

To begin with, it is interesting to note that the structure of the divergent terms in the entanglement entropy is richer than the one of the slab geometry.
The divergent piece exhibits both power-like and logarithmic terms \cite{wen}\footnote{We have some different numerical factors w.r.t. \cite{wen}.}
\be
S_{div} = \frac{(2 \pi)^4 R^3}{3G_N M_{KK}}\left[\frac12 r^2 u_{\infty}^2 - \frac32 R^3 u_{\infty} + \frac18 \frac{R^6}{r^2} \log{\frac{u_0}{u_{\infty}}}\right]\,.
\ee
In terms of field theory variables this can be rewritten as
\be
S_{div} =  \frac{8}{3^5}\frac{M_{KK}^2}{(1+\Theta^2)^2} \lambda_4 N_c^2 \left[\frac{r^2} {\epsilon^2} - \frac{27}{4\epsilon}\frac{(1+\Theta^2)}{M_{KK}^2} +\frac{81}{64} \frac{(1+\Theta^2)^2}{r^2 M_{KK}^4} \log\epsilon\right]\,,
\ee
where $\epsilon=u_0/u_{\infty}\rightarrow0$ accounts for the short distance divergence.

As in the previous section, there are two competing solutions with different topology and a phase transition between the two at a certain critical radius. 
The large $r$ behavior is dominated by a cylinder-like solution, for which the entanglement entropy reads\footnote{Again, we have some different factors w.r.t. \cite{wen}.}
\be\label{eesphere}
S_{cyl} = \frac{(2\pi)^4 R^3}{3G_N M_{KK}}\left[\frac12 r^2(u_{\infty}^2-u_0^2)+ \frac{R^3 u_0}{4}\left(-6 \frac{u_\infty}{u_0}+2+\frac{\pi}{\sqrt{3}}+\log{3}\right)\right]\,.
\ee
The first, $r^2$-dependent part is the analogous to the result for the disconnected surface in the slab geometry (\ref{eeslabdiscb}): the two expressions coincide\footnote{Modulo a factor of $\frac12 $ due to the two branches of the disconnected surface in the slab case.} if we denote the volume of the two-sphere as $V_2$.
But in the present geometry there are also subleading terms.
In particular, the $r^2$-independent term in (\ref{eesphere}), which in $(2+1)$d is by definition the (opposite of the) topological entropy, reads
\be
S_{r^2-ind} = \frac{1}{54} \frac{\lambda_4 N_c^2}{(1+\Theta^2)} \left[-\frac{6}{\epsilon}+2+\frac{\pi}{\sqrt{3}}+\log{3}\right]\,.
\ee
Again, since this is a IR quantity, it is very sensitive to the presence of the $\theta$ term.
Note that for this geometry also in the IR (``confined'') region one has $\partial S/\partial r \sim {\cal O}(N_c^2)$. 
\section{Conclusions}
In this work we have studied the $\theta$ dependence in Witten's large $N_c$ Yang-Mills model using the holographic correspondence. 

The model allows to extract the exact $\theta$ dependence of a class of interesting observables, like the vacuum energy density, the string tension, the mass of the baryon vertex, the 't Hooft loop, the confinement-deconfinement critical temperature, the mass of the $0^{++}$ glueball and the entanglement entropy. 
The model shows a common trend of all the mass scales of the theory: they get reduced by $\theta$. 
More precisely, they scale as inverse powers of $(1+\Theta^2)$ with $\Theta \sim \lambda_4 (\theta +2 \pi k)/N_c$.
Moreover, the model explicitly realizes the expected periodic structure of the $(T,\theta)$ phase diagram, with triple points where first-order transition lines meet, see figure \ref{fig2}. 
It also provides an explicit realization of oblique confinement, with the expected relations among the string tensions.

In the classical gravity regime, valid for strong 't Hooft coupling ($\lambda_4\gg1$) and large $N_c$, Witten's model consists of an $SU(N_c)$ gauge theory coupled with adjoint Kaluza-Klein massive matter fields. 
Despite this feature, we have shown that the model shares with pure Yang-Mills not only all the relevant IR features, like confinement and the occurrence of a mass gap, but also the behavior with $\theta$ of a class of relevant observables. 
The comparison in this case has been done looking at the results obtained on the lattice for small $\theta$. 
The qualitative agreement we have found in the sign of the first $\theta^2$ corrections let us hope that the holographic results, which are exact in $\theta$, can provide useful benchmarks and stimulate further lattice analysis on the subleading corrections. 
\vskip 15pt \centerline{\bf Acknowledgments} \vskip 10pt \noindent We are grateful to Ettore Vicari for motivating us to perform part of the computations presented in the paper and for illuminating discussions on the lattice results. We thank Claudio Bonati, Massimo D'Elia, Adriano Di Giacomo, Anton Faedo, Andrei Parnachev, Maurizio Piai, Erik Tonni for relevant comments and discussions. 
We thank the Galileo Galilei Institute for Theoretical Physics for the hospitality and the INFN for partial support during the completion of this work.
Finally, we thank the JHEP referee for his observations and suggestions.
\appendix
\section{The $\theta$-backreacted solution from the 1d effective action}\label{appA}
\setcounter{equation}{0}
Let us look for a regular solution to the action (\ref{action1}). As in \cite{Aharony:2006da,smearedSS}, let us consider the following ansatz for the 10d metric
\be
ds_{10}^2 = e^{2\lambda} dx_{\mu} dx^{\mu} + e^{2\tilde\lambda}dx_4^2 + l_s^2 e^{-2\varphi}d\rho^2+l_s^2 e^{2\nu}d\Omega_4^2\,,
\ee
where $\mu=0,\dots 3$ label the 4d Minkowski coordinates of the spacetime where the dual Yang-Mills field theory is defined, $\lambda,\tilde\lambda,\varphi,\nu$ are functions of the radial coordinate $\rho$ and the $x_4$-coordinate is compactified on a circle of length $\beta_4 = 2\pi M_{KK}^{-1}$.
The function $\varphi$ is related to the dilaton $\phi$ through the defining equation
\be
\varphi = 2\phi - 4\lambda - \tilde\lambda - 4\nu\,.
\ee
The transverse part of the metric includes a four-sphere $S^4$. The flux of $F_4$ through $S^4$ is quantized and proportional to $N_c$, see eq. (\ref{f42}). 

We will consider the following ansatz for $F_2$
\be
F_2 = dC_1 = \dot h\, dx^4\wedge d\rho\,,
\ee
where $h$ is a function of the radial variable $\rho$ and the dot stands for derivative with respect to $\rho$.  

Implementation of the ansatz above gives the following 1d action
\bea
S &=& {\cal V}\int d\rho \left[-4{\dot\lambda}^2 - {\dot{\tilde\lambda}}^2 - 4{\dot\nu}^2 + {\dot\varphi}^2 -\frac12 e^{4\lambda-\tilde\lambda+\varphi+4\nu}{\dot h}^2 +V\right]\,,\nonumber \\
V&=& 12 e^{-2\nu-2\varphi}-Q_c^2 e^{4\lambda+\tilde\lambda-4\nu-\varphi}\,,
\eea
which has to be supported by the zero-energy constraint
\be
-4{\dot\lambda}^2 - {\dot{\tilde\lambda}}^2 - 4{\dot\nu}^2 + {\dot\varphi}^2 -\frac12 e^{4\lambda-\tilde\lambda+\varphi+4\nu}{\dot h}^2 - V=0\,.
\label{ze}
\ee
 Above we have defined (using $R^3 = \pi g_s N_c l_s^3$)
\be
Q_c = \frac{3}{\sqrt{2}g_s}\frac{R^3}{l_s^3}= \frac{3\pi N_c}{\sqrt{2}}\,,
\ee
as the constant arising from the quantized $F_4$ flux through $S^4$ and
\be
{\cal V}= \frac{1}{2k_0^2} V_{1,3} V_{S^4}\frac{2\pi}{M_{KK}}l_s^3\,,
\ee
where $V_{1,3}$ is the infinite 4d space-time volume and $V_{S^4}= 8\pi^2/3$ is the volume of $S^4$.
\subsection{Equations of motion and general solutions}
 The equation of motion for the field $h$ implies that
 \be
 \dot h = - q_{\theta}e^{-4\lambda+\tilde\lambda-\varphi-4\nu} = -q_{\theta} e^{2\tilde\lambda-2\phi}\,,
 \label{eqf}
 \ee
 where $q_{\theta}$ is a constant which we will relate to the Yang-Mills $\theta$ angle. The equations of motion for the remaining fields, which we re-arrange so that the dilaton $\phi$ appears instead of $\varphi$, read
\bea
&&\ddot{\lambda} -\frac{Q_c^2}{2}e^{8\lambda+2\tilde\lambda-2\phi}=\frac{q_{\theta}^2}{4}e^{2\tilde\lambda-2\phi}\,,\nonumber \\
&&\ddot{\tilde\lambda} -\frac{Q_c^2}{2}e^{8\lambda+2\tilde\lambda-2\phi}= - \frac{q_{\theta}^2}{4}e^{2\tilde\lambda-2\phi}\,, \nonumber \\
&&\ddot{\phi} -\frac{Q_c^2}{2}e^{8\lambda+2\tilde\lambda-2\phi}=\frac34 q_{\theta}^2 e^{2\tilde\lambda-2\phi} \,, \nonumber \\
&&\ddot{\nu} +\frac{Q_c^2}{2}e^{8\lambda+2\tilde\lambda-2\phi}-3e^{8\lambda+2\tilde\lambda-4\phi+6\nu}=\frac{q_{\theta}^2}{4}e^{2\tilde\lambda-2\phi} \,.
\eea  
Interestingly, defining
\be 
\gamma = 8\lambda+2\tilde\lambda-2\phi\,,\quad p = 8\lambda + 2\tilde\lambda -4\phi + 6\nu\,,\quad \chi = 2(\tilde\lambda-\phi)\,,
\label{hpchi}
\ee
we get the following decoupled differential equations
\be
\ddot \gamma = 4 Q_c^2 e^{\gamma}\,,\quad\ddot p=18\,e^p\,, \quad \ddot \chi = -2 q_{\theta}^2 e^{\chi}\,,
\label{eqch}
\ee
which can be easily solved in terms of elementary functions
\bea
\gamma &=& -2\log[a_1 - e^{-a_2\rho}] - a_2\rho + \log\left[\frac{a_1a_2^2}{2Q_c^2}\right]\,,\nonumber\\
p &=& -2\log[a_3 - e^{-a_4\rho}] - a_4\rho + \log\left[\frac{a_3a_4^2}{9}\right]\,,\nonumber\\
\chi &=& -2\log[a_5 + e^{-a_6\rho}] - a_6\rho + \log\left[\frac{a_5a_6^2}{q_{\theta}^2}\right]\,.
\eea
These in turn imply that
\be
\nu = -\frac18 (\gamma+\chi) + \frac{p}{6} + a_7 + a_8\rho\,.
\ee
We have 8 integration constants $a_1,...,a_8$ to be fixed imposing IR regularity and the constraint (\ref{ze}) to begin with. 

Another requirement is that when $q_{\theta}=0$ the solutions read \cite{Aharony:2006da} \bea\label{solwitten}
&&\lambda_0 (\rho) =  -\frac{1}{4}\log\left[1-e^{-3a\rho}\right] + \frac{3}{4} \log{\frac{u_0}{R}}\,,\nonumber \\
&& \tilde\lambda_0(\rho) =  -\frac{1}{4}\log\left[1-e^{-3a\rho}\right] -\frac{3}{2}a\rho +  \frac{3}{4} \log{\frac{u_0}{R}}\,,\nonumber \\
&& \phi_0 (\rho) =  -\frac{1}{4}\log\left[1-e^{-3a\rho}\right] + \frac{3}{4} \log{\frac{u_0}{R}}+\log g_s\,,\nonumber \\
&& \nu_0(\rho) = -\frac{1}{12}\log\left[1-e^{-3a\rho}\right] + \frac{1}{4} \log{\frac{u_0}{R}}+\log\frac{R}{l_s}\,,
\eea
where we have introduced the parameter
\be\label{defa}
a\equiv \frac{\sqrt{2} Q_c u_0^3}{3R^3 g_s}= \frac{u_0^3}{l_s^3 g_s^2}\,,
\ee
where $u_0$ is the minimal value of the radial variable $u$ used in the main body of this paper.
As in the $q_{\theta}=0$ case, the variable $\rho$ is related to the radial variable $u$ by
\be\label{rhou}
e^{-3a\rho} = 1-\frac{u_0^3}{u^3}\,,
\ee
so that $\rho\rightarrow 0$ (resp. $\rho\rightarrow\infty$) when $u\rightarrow\infty$ (resp. $u\rightarrow u_0$).
\subsection{Fixing some of the integration constants}
From the expressions above we can deduce the following constraints. First of all, we look for a regular confining solution in the IR. Requiring that $e^{2\lambda}$ goes to a constant in the IR amounts to choose
\be
a_6=a_2\,.
\ee
If in turn we require that, as in the $\theta=0$ case, $e^{2\lambda}\rightarrow\infty$ in the UV we need to require
\be
a_1=1\,.
\ee
Moreover, requiring $e^{2\nu}$ to go to a constant in the IR and to diverge in the UV we find
\be
a_8=-\frac14 a_2+\frac16 a_4\,,\quad a_3=1\,.
\ee
Requiring $e^{2\tilde\lambda}$ to go to zero in the IR (so that the $x_4$ circle shrinks as in the $\theta=0$ case) we find the condition
\be
a_4 < 2 a_2\,.
\ee
The zero-energy constraint (\ref{ze}) in turn fixes (once all the above conditions are satisfied)
\be
a_4 = a_2\,.
\ee
Once all the conditions above are fulfilled the $(x_4,r)$ subspace is regular in the deep IR if 
\be
\frac{3^{2/3} a_2^{4/3} e^{4 a_7}}{\sqrt{2}\sqrt{a_5}l_s^2 Q_c q_{\theta}}=\frac49 M_{KK}^2\,.
\label{mkka}
\ee
\subsection{A particular solution}
Since the equations for $\gamma$ and $p$ are the same in form as in the $q_{\theta}=0$ case, a first simple choice for the yet-to-be fixed integration constants reads
\be
a_2=3a\,,
\ee
so that $\gamma=\gamma_0$ and $p=p_0$. Using the above choice, we get
\be
a_8 = -\frac{a}{4}\,.
\ee
Notice that with these choices
\be
e^{\chi(\rho)} = \frac{a^2 a_5 e^{3a\rho}}{q_{\theta}^2 (1+a_5 e^{3a\rho})^2}\,.
\ee
Since this is identified with the running gauge coupling (see e.g. \cite{Bigazzi:2004ze}) , and we expect $\theta$ not to modify it in the deep UV, we require that
\be
e^{\chi(0)} =e^{ \chi_0(0)}=\frac{1}{g^2_s}\,,
\ee
which sets
\be
a_5 = -1 +\frac{9a^2g_s^2}{2q^2_{\theta}}\left(1+\sqrt{1-\frac{4q_{\theta}^2}{9a^2 g_s^2}}\right)\,,
\ee
where we have selected the value such that at $q_{\theta}=0$, $\chi=\chi_0$. When $q_{\theta}\rightarrow0$ we have, in fact
\be
a_5\rightarrow \frac{9 a^2 g_s^2}{q_{\theta}^2}\,,
\ee
so that $\chi\rightarrow\chi_0$ consistently.

If, from (\ref{mkka}), we want to get the usual relation between $u_0$ and $M_{KK}$ in the $q_{\theta}\rightarrow0$ limit, and if we assume $a_7$ to be independent from $q_{\theta}$, we can also fix
\be
a_7 = \frac16 \log g_s\,.
\ee
As for the RR potential, we get
\be
C_1 = h\, dx^4 =\frac{3a}{q_{\theta}} \left[-\frac{a_5}{a_5 + e^{-3a\rho}}+1\right] dx^4\,,
\ee
where (in order to get a regular solution) we have imposed it to vanish in the deep IR ($\rho\rightarrow\infty$) where the the $x_4$ circle shrinks. The Yang-Mills $\theta$ angle is identified (mod $2\pi$) with the integral of $C_1$ along the $S_{x_4}$ circle at $\rho\rightarrow0$
\be
\theta +2\pi k=\frac{1}{l_s} \int_{S_{x_4}} C_1 = \frac{2\pi}{M_{KK}}\frac{3a}{q_{\theta}\,l_s (1+a_5)} = \frac{2\pi}{M_{KK}}\frac{2q_{\theta}}{3a\, l_s g_s^2}\frac{1}{1+\sqrt{1-\frac{4q_{\theta}^2}{9a^2g_s^2}}}\,.
\label{thetaq}
\ee
In the expression above $k\in\mathbb{Z}$.
When $q_{\theta}\sim 0$ we get (setting $k=0$)
\be
\theta\rightarrow\frac{2\pi}{M_{KK}}\frac{q_{\theta}}{3a\, l_s g_s^2}= \beta_4 \frac{q_{\theta}}{3a\, l_s g_s^2}\,,
\ee
and so
\be
C_1\rightarrow l_s \frac{\theta}{\beta_4} e^{-3a \rho} = l_s \frac{\theta}{\beta_4}\left(1-\frac{u_0^3}{u^3}\right)\,,
\ee
which is (re-written in terms of the original radial variable $u$) the known expression one gets (for small $\theta$) at first order in $\theta$ \cite{Witten:1998uka}. 

Notice that taking the opposite limit $q_{\theta}\rightarrow\infty$ is not allowed as we must have
\be
|q_{\theta}|\le\frac32 a g_s\,.
\label{boundq}
\ee
When $q_{\theta}\rightarrow \pm(3/2) a g_s$ we get (setting $k=0$) $\theta\rightarrow\pm\beta_4/(l_s g_s)=\pm4\pi^2 N_c/\lambda_4$, where $\lambda_4=4\pi^2 g_s N_c l_s /\beta_4$ is the 't Hooft coupling at the KK scale $M_{KK}$.\footnote{Notice that when $q_{\theta}\rightarrow \pm(3/2) a g_s$, $a_5\rightarrow1$.}
On the other hand, the validity of the gravity approximation requires $\theta \ll 4\pi^2 N_c / \lambda_4^{3/4}$ \cite{dubovsky}.
Thus, while $\theta$ can be of order $N_c$ as expected, $\theta/N_c$ must be small.
\subsection{The particular solution in the $u$-variable}
Let us now see how the particular solution found above reads in terms of the radial coordinate $u$ defined, as in the $q_{\theta}=0$ case, in (\ref{rhou}). Let us also define
\be
w\equiv\frac23\frac{q_{\theta}}{a\, g_s}\,,\quad {\rm so\, that}\,\,|w|\le1\,.
\ee
Using ({\ref{thetaq}) we get \be
w = \frac{2 \Theta}{1+\Theta^2}\,,
\ee
where, for any $k$
\be
\Theta\equiv\frac{\lambda_4}{4\pi^2}\frac{\theta+2\pi k}{N_c}\,.
\ee
Notice that the bound (\ref{boundq}) is automatically satisfied by any $\Theta$. 

Using these redefinitions, the particular solution found above turns out to precisely coincide with that 
written in the main body of the paper and already found in \cite{pasquinucci, dubovsky}.
 
\end{document}